\documentstyle[a4,12pt]{article}

\def\APP{{\em Acta Physica Polonica} }

\def\NPB{{\em Nucl. Phys.} B}
\def\PLB{{\em Phys. Lett.}  B}
\def\PRL{{\em Phys. Rev. Lett.}}
\def\PRD{{\em Phys. Rev.} D}
\def\ZPC{{\em Z. Phys.} C}

\def\EPJC{{\em Eur. Phys. Journal} C}
\def\ARNS{{Annu. Rev. Nucl. Sci.}}

\def\ra{\rightarrow}

\def\tb{\tan \beta}
\def\ch{H^{\pm}}
\def\sab{\sin(\beta-\alpha)}

\def\be{\begin{equation}}
\def\ee{\end{equation}}
\def\bea{\begin{eqnarray}}
\def\eea{\end{eqnarray}}
\def\bi{\bibitem}
\begin{document}
\begin{flushright}
DESY 98-177\\
IFT 98/8\\
FERMILAB-PUB-98/353-T\\[1.5ex]
{\large \bf hep-ph/9811256} \\
% version 6.2p \\
\today\\
\end{flushright}

\begin{center}
{\bf \Large PROCESS $Z{\rightarrow}h(A)+\gamma$ IN THE 2HDM \\
                       AND THE EXPERIMENTAL CONSTRAINTS FROM LEP}
\end{center}

\vskip 0.4cm
\centerline{MARIA KRAWCZYK
%\footnote{also at DESY, Hamburg, Germany} 
            and JAN \.ZOCHOWSKI}

\centerline{{\it Institute of Theoretical Physics, University of Warsaw, }}

\centerline{\it Warsaw, 00-681, Poland}
\vskip 0.5cm
\centerline{and}
\vskip 0.5cm
\centerline{PETER M\"{A}TTIG}

\centerline{\it Weizmann Institute}

\centerline{\it Rehovot, Israel}

% \centerline{$^*$ also at DESY, Hamburg, Germany} 

\begin{abstract}

The one-loop  branching ratios for the process
$Z{\rightarrow}h(A)+\gamma$ 
are calculated in  the general Two Higgs Doublet Model (Model II)
taking into account existing constraints on the model parameters.
For Higgs boson masses below 50 GeV and 
$\tb $ ${\cal O}(1-10)$ the fraction of such $Z$ decays are 
at the level of $10^{-7}$,
but can be 
significantly stronger for very low or high $\tb $, where
the dependence of these results on other model parameters like
$\sin (\beta - \alpha )$ and the mass of the charged Higgs boson
is found to be of little importance.
The results are compared to  the 
LEP measurements, which are sensitive to branching ratios of 
$Z{\rightarrow}h(A)+\gamma$
of the order 
$10^{-5}$ for masses $\ge $ 20 GeV, 
but approach $10^{-6}$ for low masses. 
Relating the expectation to the experimental limits,
constraints on the parameter space of the 2HDM
are derived.
\end{abstract}

\section {Introduction}

The Two Higgs Doublet extension of the Standard Model 
 leads to five physical Higgs particles: two neutral scalars $h$, $H$ 
 (with the mass relation $M_H>M_h$),
 one pseudoscalar $A$
and two charged particles $H^{\pm}$.  In case of CP conservation, 
their interaction with fermions 
and gauge bosons is characterized by only 
two additional parameters $\alpha$ and $\beta $, describing  the 
mixing within the neutral scalar system and the ratio of the
vacuum expectation values, respectively 
\cite{hunter}. The Higgs bosons couple also to themselves and 
this self-coupling requires an extra parameter $\lambda_5$.
A lot of attention has been devoted to Two Higgs Doublet models embedded
in the Minimal Supersymmetric Model (MSSM).
Here strong relations between the various masses and also
with the parameters $\alpha$ and $\beta $ exist, 
such that at the tree level there are only two independent parameters and 
stringent
experimental limits on Higgs masses can be set.
In this paper we discuss the CP-conserving Two Higgs Doublet Model II,
denoted 2HDM, which has a Higgs sector as in MSSM,
but where these relations do not exist
and the masses of the Higgs particles are very weakly
constrained.

In this model
the couplings of the pseudoscalar $A$ to fermions 
are given in terms of $\beta$.
Couplings to gauge bosons $AWW$ and $AZZ$ are forbidden.
The couplings of the scalar $h$
to fermions (and gauge bosons)  
depend in addition on $\alpha $.
For example, the coupling $hZZ$ boson contains
factors $\sin(\beta-\alpha)$, and is thus
suppressed for $\alpha\sim \beta$.
Theoretically the allowed ranges of $\alpha $ and $\beta $ 
are only  constrained 
through the requirement of
the perturbativity of calculations which suggests
$\tb$ to be between $\sim$ 0.1  and 200-300 \cite{lim}. 

Important constrains on the neutral sector
of the general 2HDM are due to searches for Higgs boson production in
$Z$ decays.
~From the absence of evidence for the Higgs-strahlung
process ($Z \rightarrow Z +  h$) limits on the
$\sin^2(\beta-\alpha)$ as a function of $M_h$ can be inferred.
At least for up to $M_h \ \sim $ 50 GeV they imply 
$\mid \sin^2 (\beta-\alpha) \mid \ < $ 0.1 and thus
$\alpha \sim \beta $ \cite{sin,mhmasin,mhmaOPAL,bib-sinL3}.
Complementary to the Higgs-strahlung, the 
decay  
$Z \rightarrow   h + A$ is proportional to $\cos^2(\beta-\alpha)$.
Also for the Higgs pair production process no evidence has been found.
Combining
the sensitivities reached for these two production mechanisms,
one can derive a limit on the sum of the two Higgs masses:
$M_h + M_A $ has to be larger than about 
50 GeV~\cite{bib-Desch,mhma,mhmasin,mhmaOPAL}.
For the Model II, if embedded in supersymmetry, the same measurements
exclude both a pseudoscalar or scalar neutral Higgs boson of 
less than $\sim $ 77 GeV for $\tb > 0.8$ \cite{bib-Desch}. 
However, in the 2HDM, because of 
the absence of relations between masses and between other parameters,
no limits on the masses of individual Higgs bosons can be set:
even a very light neutral Higgs particle  is not excluded.

Another potential production mechanism for a Higgs particle 
at LEP is a Yukawa process where  
Higgs particles are radiated from heavy fermions, namely 
$Z \rightarrow b\bar {b}h(A)$, 
$Z \rightarrow \tau^+\tau^- h(A)$.
As yet measurements \cite{bib-al2HDM} have only been interpreted 
in terms of limits
on $\tb $ and $M_A$.
%Also in this case no absolute limit can be set for Higgs masses.
For $\tb >$ 25, only Higgs boson 
masses of less than $\sim $ 2 GeV are excluded
by these data  and much 
larger values for $\tb $ are allowed for
higher Higgs boson masses.
For small masses an interpretation of the data in terms of 
$M_h$ production would
yield stronger limits on $\tb $ (see \cite{bib-mariajan}).
Some further constraints on neutral Higgs bosons 
are obtained from non-LEP experiments.  
The present data of 
$g-2$ for muons limit
the allowed $\tb$ for the pseudoscalar or scalar mass below  2 GeV 
to values of 4 at $M_h$=0.1 GeV~\cite{g2},
for higher masses the limits on $\tb $ are 
weaker than those from the Yukawa process 
\cite{bib-al2HDM}.
The measurement of the Wilczek process 
$J/\psi, \Upsilon \rightarrow h(A) + \gamma $
points to possible constraints for the $M_{h(A)}$ 
below 10 GeV \cite{wilk,mk},
unfortunately  the interpretation suffers both from theoretical uncertainties
and lack of experimental considerations of some aspects of the 2HDM. 
In conclusion, only very weak limits exist for this rather
simple extension of the Higgs sector.
It is therefore important to search for additional 
relevant experimental data,
particularly if it constrains the masses of $h$ and $A$ bosons independently.

In this paper we study the 2HDM contribution
to  
$Br(Z{\rightarrow}h(A)+\gamma)=\Gamma (Z{\rightarrow}h(A)+\gamma)
/\Gamma(Z\ra all)$, where the $\Gamma $ denote the partial, respectively, 
total width of the $Z$, and compare these to experimental data. 
For the theoretical evaluation we 
take into account existing LEP limits on the model parameters.
In detail, the calculations include the following results, which
are all valid at 95$\% $ confidence:
\begin{enumerate}

\item The exclusion on $\sin^2(\beta-\alpha)$ 
      for $M_h$ smaller than 60  GeV \cite{bib-sinL3}~\footnote{
      Recently limits on $\sin^2(\beta-\alpha)$ became available from other
      experiments as well~\cite{sin,mhmasin,mhmaOPAL} which in addition
      extend towards higher Higgs masses. Because of lack of detailed
      information we refrain from combining these individual limits.
      In addition, because of the experimental sensitivity the constraints
      on $\sin^2(\beta-\alpha)$ for high masses would not add significantly
      to our conclusion.}.

\item The limit on $\tb$ for $M_A$ smaller then 40 GeV 
       \cite{bib-al2HDM}.

\item  The excluded region of 
       $M_h$ versus $M_A$ \cite{mhma}.

\end{enumerate}
In addition the result from the NLO analysis of the $b\ra s + \gamma$ process 
is invoked:
\begin{itemize}
\item[4.]  The mass of the charged Higgs boson 
should be larger than 330-350 GeV \cite{bsg}
\footnote{This limit is based on the published CLEO data 
\cite{bib-CLEObsgapub}.
Recently the ALEPH collaboration has published a new analysis 
\cite{bib-ALEPHbsgapub}
and CLEO released new preliminary results \cite{bib-CLEObsgaprel}.
The results tend to relax the limits on the charged Higgs boson,
a new theoretical analysis leads to a lower limit of 165 GeV \cite{greub}.
Note that also the Tevatron searches for $t\rightarrow H^{\pm} X$ 
\cite{CDFH+} lead
to constraints only in a limited region of parameter space in the 2HDM
\cite{CDFH+_HO_corrections}. See also \cite{bor}.}. 
Alternatively we also consider the mass limit for a charged Higgs 
from the direct search at LEP yielding 
to $M_{H^{\pm}}>54.5$ GeV \cite{bib-LEPchargedH}{\footnote{
Preliminary results from LEP data at 183 GeV set limits of up to 59 
GeV~\cite{bib-Desch}.}}.  

\end{itemize}

Relevant experimental results from LEP1 on the search for 
$Z{\rightarrow}h+\gamma$ have been published by all four LEP experiments
\cite{bib-ALEPH,bib-DELPHI,bib-L3,bib-OPAL}.
The decay modes considered include 
$h\rightarrow b\bar{b}, \ \tau ^+\tau ^-$ and inclusive
hadrons, which are independent of quark flavours and applicable also 
to decays into a pair of gluons.
The mass range covered is between 5 and 85 GeV.
The experiments are typically sensitive to branching ratios 
$Br(Z\rightarrow h+\gamma )\cdot Br(h\rightarrow X)$ of 
${\cal O}(10^{-5})$ but approach 10$^{-6}$ for low $M_h$ and 
$X \ = \ $ hadrons or $\tau ^+ \tau ^- $.
Note that since the angular distribution for the $h$ and $A$
final state are identical up to a normalization factor \cite{shimada}
these experimental results should also hold for the pseudoscalar $A$.

In this paper we will first address the theoretical aspects of this process,
the production rates and  decay modes for a neutral Higgs boson
as a function of its mass
and for various values of $\tb $.
The dependence on the charged Higgs boson mass and their
coupling to the neutral scalar is also discussed.
We then summarise the experimental situation and finally conclude
on its relevance for constraining the parameter space of the 2HDM.

In this paper we restrict ourselves to the one-loop
contributions to decays of on-shell $Z$'s.
This decay in the SM were studied in \cite{gun}a-c 
\footnote{Note that the QCD corrections 
 were calculated in \cite{spira}, they were found to be small.}, in 
\cite{dju} the SM, 2HDM
and MSSM was also discussed.
A more general theoretical analysis of $h(A)+\gamma $ production 
which also addresses energies
above the $Z$ peak can be found, for example in~\cite{gun}b-e, \cite{rosiek}. 

% =====================================================================

\section{The process $Z{\rightarrow}h+\gamma$ in the Standard Model}

As a reference we summarise the theoretical results on the 
$Z$ decay into $h + ~\gamma $ within the Standard Model. 
Here the process would be mediated by 
$W$ and fermion loops \cite{hunter}, \cite{gun}a-c, \cite{dju}.
In Fig.~\ref{fig:SMprod}a
the branching ratio 
$Br(Z{\rightarrow}h+\gamma)$
is shown as a function of the
scalar Higgs boson mass.
Also   the individual contributions
to the branching ratios are displayed.
As can be seen the
$W$-loop contributes almost exclusively to this process. 
% (two   highest dotted lines correspond to the  $W$ and to the 
% slightly lower total contribution).
Note that there is a relative minus sign between
the $W$- and fermion terms.

As can be seen from Fig.~\ref{fig:SMprod}a, the Standard Model branching ratio
is below 5$\cdot $10$^{-6}$ in the whole mass range 
and thus beyond the experimental sensitivity.
Anyhow, a Standard Model Higgs of mass less than 89.9 GeV has been
excluded
from searches at LEP for the Higgs-strahlung \cite{lepwork}. 
As discussed in the introduction, these limits do not apply
in the 2HDM.
Here the $Z$ decay into $h(A)+ \gamma $ can be in principle 
stronger and may
provide the most prominent signal for Higgs production
for some regions in parameter space.

% ******************************************************************

\section{The process $Z{\rightarrow}h(A)+\gamma$  
in the 2HDM}
We start our analysis of $Z$ decays into photons and $h(A)$ within
the 2HDM (see also \cite{hunter} and especially \cite{dju}) by
listing the Higgs couplings to quarks and gauge bosons 
in a form which will make our discussion more transparent~\cite{haba}.
For the coupling to fermions 
the SM factor $(-igm_f/2M_W)$ is modified by factors which differ for the two
fermion isospins, 
 for example for bottom and top quarks: 
\bea
hb {\bar b}:{\hspace{0.4cm}} {{-\sin\alpha}\over{\cos\beta}}=
\sin(\beta-\alpha)-\tb \cos(\beta-\alpha)\\
ht {\bar t} :{\hspace{0.5cm}} {{\cos\alpha}\over{\sin\beta}}=
\sin(\beta-\alpha)
+{{1}\over{\tb}} 
\cos(\beta-\alpha)
\eea
The $h$ couples to $ZZ$ with a SM factor
 ($ig {{M_Z} / {\cos\theta_W}} ~g^{\mu \nu}$) times
\be
hZZ: {\hspace {0.5cm}} \sin(\beta-\alpha). 
\ee
For our further considerations two extreme cases of parameters
are of interest:
\begin{itemize}

\item case A 
$$\cos(\beta-\alpha)=0 ~({\rm  equivalently \hspace{0.2cm}} 
\sin(\beta-\alpha)= +1 
)\footnote{For the purpose of our analysis the other sign will not
be considered.}$$
which corresponds to the SM case, since for both 
the $hb{\bar b}$ and $ht{\bar t}$ 
as well as for the $hZZ$ couplings the factors of eqs.~1-3 are unity.  
Note that there is a relative minus sign between fermionic and
gauge coupling.
There is no dependence on $\tb$.

\item case B   
$$\sin(\beta-\alpha)=0 ~({\rm equivalently \hspace{0.2cm} }
\cos(\beta-\alpha)=+1~ {\rm or }\hspace{0.2cm} \alpha=\beta )$$
which leads to a scenario that is totally 
different from the Standard Model one.  
Here  the $hZZ$ coupling disappears,
moreover $hb{\bar b}$ and $ht{\bar t}$ couplings 
have opposite  signs, independent of
whether we choose $\cos(\beta-\alpha)$=1 or -1.
So even $\tb$ =1 does not necessarily correspond to the SM prediction
although, for special cases, i.e. if one contribution dominates, 
it looks like the 
Standard Model.
Note that for a large value of $\tb$ the Higgs scalar $h$ 
may have a larger coupling to the bottom quark, than to top, despite the 
larger top quark mass.

\end{itemize}
For the coupling of the pseudoscalar $A$ to fermions 
the corresponding factors are 
\bea
Ab {\bar b} : {\hspace{0.3cm}} -i\gamma_5\tb~~ \\
At {\bar t} : {\hspace{0.5cm}} -i\gamma_5{{1}\over{\tb}}. 
\eea
The $AZZ$, $AWW$ couplings are absent in the considered model \cite{hunter}.

\subsection{$Z{\rightarrow}h+\gamma$}
In the 2HDM \cite{hunter,dju}
$W$, charged leptons or down-type quarks,
and up-type quarks contribute to the matrix element for the 
$Z\rightarrow h + \gamma $ decay with factors 
given above. 
An additional contribution, not existent in the Standard Model, 
is due to loops involving charged Higgs scalars.
However, for masses of $M_{H^+} > $ 330 GeV, as required by some  
$b\rightarrow s + \gamma $ analysis, 
it is negligible.
As will be discussed in Sec.~5, this does not change for 
lower masses of $H^{\pm}$ in an important way.
%%%%%%%%%%%%%%%%%%%%%%%%%%%%%%%%%%%%%%%%%%%%%%%%%%%%%%%%%%%%%%%%%%%%%%%%

\begin{figure}[ht]
\vskip 4.in \relax\noindent\hskip -1.in
                \relax{\includegraphics{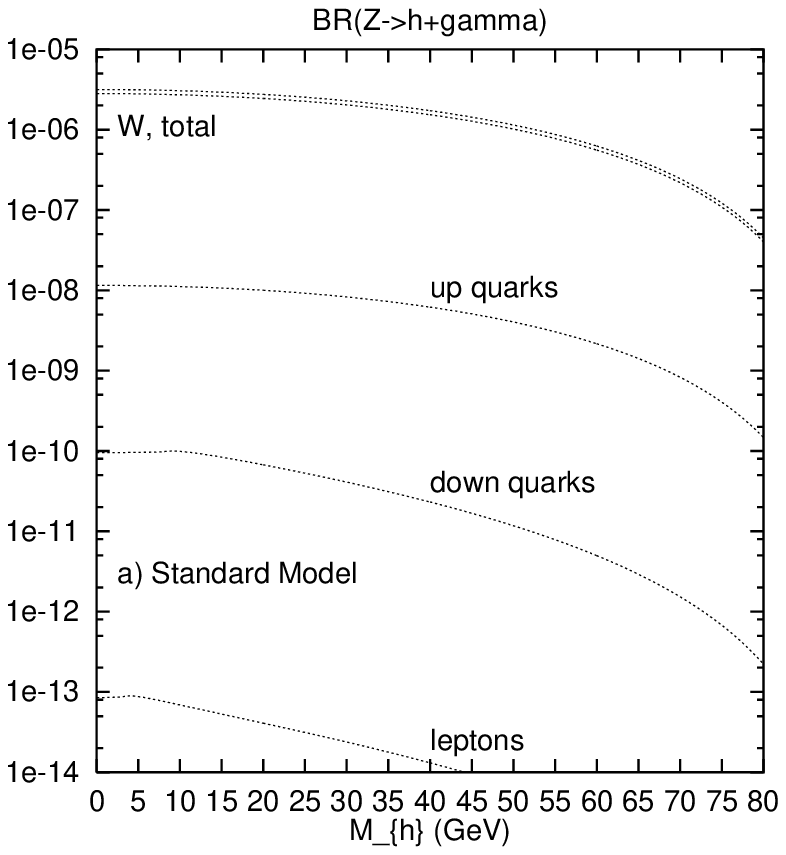}}
                  \relax\noindent\hskip 3.0in
                \relax{\includegraphics{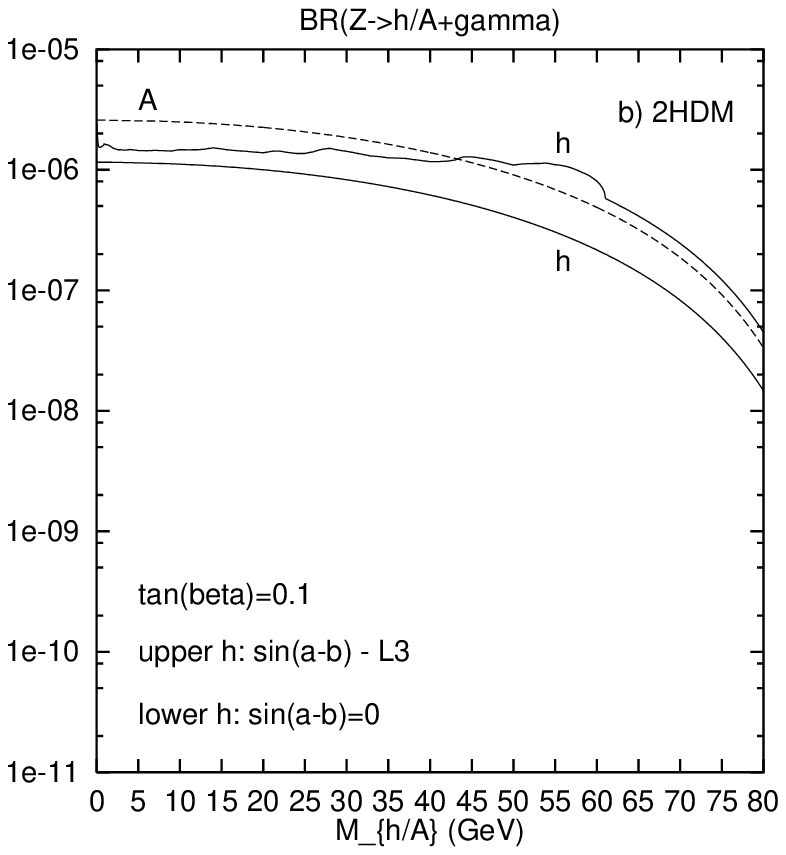}}
\vskip 3.5 in \relax\noindent\hskip -1. in
                \relax{\includegraphics{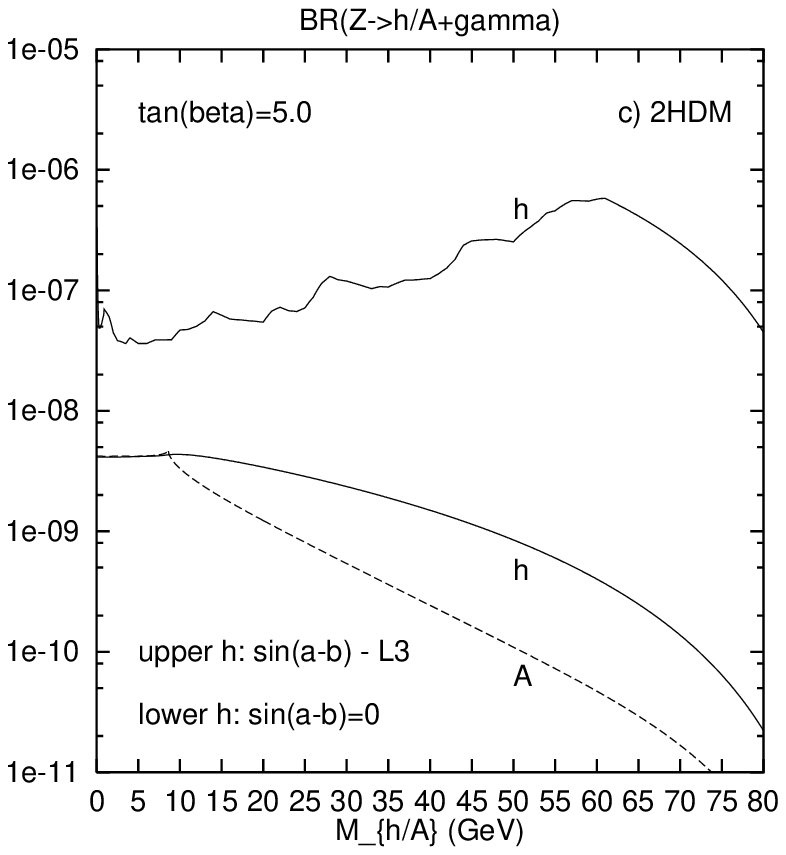}}
                 \relax\noindent\hskip 3.0 in
                \relax{\includegraphics{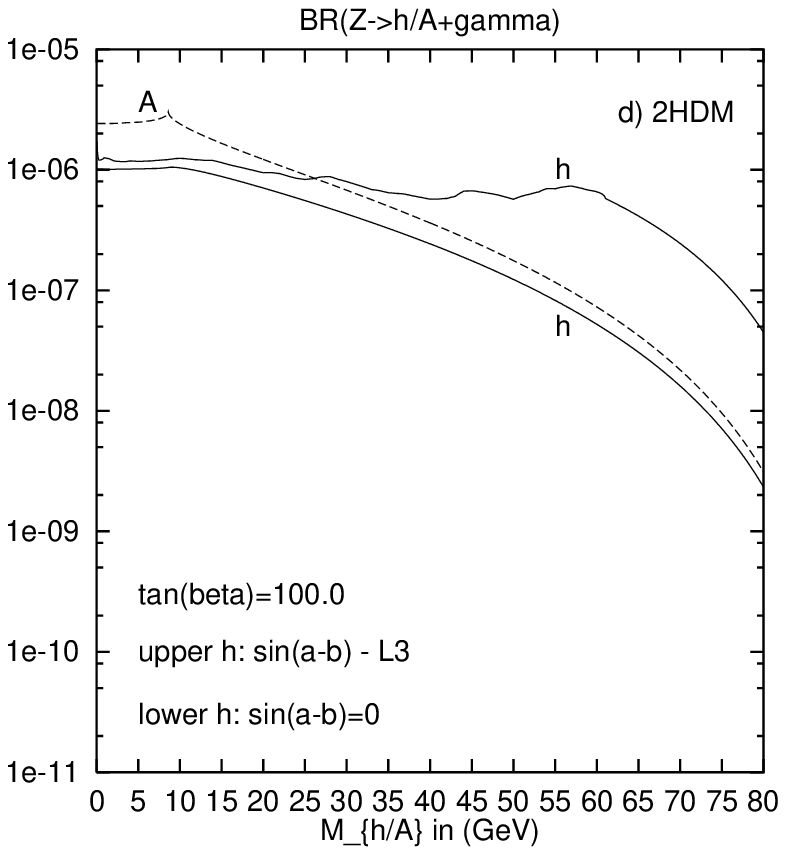}}              
\vspace{-10.5ex}
\baselineskip 0.4cm
\caption{\sl a)The scalar production in  SM (dotted lines)
- W and total, up-type quarks, down-type quarks, leptons contributions. 
b,c,d) The 
production of a scalar (solid line, $h$) and pseudoscalar (dashed line, $A$)
in the 2HDM  for $\tb$ = 0.1, 5, 100,
respectively. 
Limits on 
$\sin^2(\beta-\alpha)$ are included for upper solid curves 
and for lower solid curves $\sin(\beta-\alpha)$=0 is assumed; 
$M_{H^{\pm}}$ is set to 330 GeV.
}
\label{fig:SMprod}
\end{figure}

The branching ratios $Br(Z\rightarrow h +\gamma )$ in the 2HDM 
are presented in  Fig.~\ref{fig:SMprod}b,c,d 
for low, medium and high values of $\tb $. 
The two solid curves for each $\tb $ 
correspond to the cases of $\sin(\beta-\alpha)$=0
and of the maximum allowed value of $\sin(\beta-\alpha)$ from \cite{
bib-sinL3}. 
% for each $M_h$, lower and upper curves, respectively.
The experimental constraints on $\sin^2(\beta-\alpha)$
lead
to the wiggles in the upper curves.
The possible range 
of $h$ production in the 2HDM
for the masses $M_h$ and $\tb $ shown in Fig.~\ref{fig:SMprod}b,c,d 
is bounded by 
the two corresponding solid curves. 

For $M_h <$ 60 GeV, where the experimental constraint on 
$\sin^2(\beta-\alpha)$ \cite{bib-sinL3} is relevant, the branching ratio
increases with increasing $\tb $ for $\tb$ larger than $\sim$ 5
(see also figures discussed in Sec. 5,6).
%[{\bf True? Analysis for other $\tb $ existing?}].
%YES !!! (see the discussion on the $\tb$ dependence in Sec.6 and 7).
For $\tb $ of ${\cal O}(1-10)$ the decay fraction is significantly below
the expected yield for a Standard Model Higgs.
This is because of
the large 
suppression of the $W$ contribution for the small
% of the contribution from W's due to the experimental limit on
$\sin^2(\beta-\alpha)$ allowed by experiments.
Only for very high $\tb $
the loop of bottom quarks,
which contributes with $(\sin \alpha /\cos \beta )^2\ \sim \ \tan ^2 \beta$,
dominates such that
branching ratios comparable to the Standard Model ones are reached.
In contrast the top quark loop contributes only by 
$1/\tan ^2 \beta$ and is therefore negligible.

A large rate can be also obtained for very small $\tb$, 
see Figs.~\ref{fig:SMprod}b and figures discussed in Secs. 5, 6.
% ~\ref{fig:lowtb}.
Here the roles of $t$ and $b$ quarks  are reversed.

For $M_h >$ 60 GeV no relevant constraint exists on 
$\sin^2(\beta-\alpha)$ in \cite{bib-sinL3}
and $\sin^2(\beta-\alpha)$=1 (case A above) was assumed. 
As discussed above, this implies  
the same coupling of the Higgs boson to fermions and gauge bosons 
as in the Standard Model.

\subsection{ $Z{\rightarrow}A+\gamma$}

In the considered  2HDM with CP conservation,
because of the forbidden $AWW$ 
and $AH^+H^-$ couplings, 
%negligible, 
the $Z\rightarrow A+\gamma $ decay
is mediated only by fermions
\cite{hunter,dju}.
Charged leptons  and down-type  quarks  (up-type quarks)
contribute  to the branching ratio with the factors, 
relative to the SM case,
of $\tan ^2\beta$ ($\tan^{-2}\beta$) independent of $\alpha $.
Thus down-type quarks dominate for large $\tb $ whereas up-type quarks
dominate for $\tb \ll $1. 

The results for corresponding  $\tb$=0.1,~5 and 100  are presented in 
Fig.~\ref{fig:SMprod}b,c,d
together with the results for scalar boson production, 
see also figures discussed in Secs. 5, 6.
The branching fraction $Br(Z\rightarrow A+\gamma )$ is 
larger than the one for scalars for masses of up to 30-40 GeV
for $\tb$=0.1 and 100. For the intermediate $\tb$
 the pseudoscalar production is lower than for the scalar.
The $\tb$ dependence will be discussed further in Secs. 5 and 6.
 
Given the strongly decreasing production yield for higher masses, we will
limit the following discussion to $M_{h,A} \le $ 40 GeV.
Note that for this mass range the experimental constraints on
$\sin ^2 (\beta -\alpha )$ are strong  and will always be taken 
into account in the following discussion.

% =====================================================================

\section{Decay modes in the 2HDM}

\begin{figure}[ht]
\vskip 3.5in \relax\noindent\hskip -1.in
                \relax{\includegraphics{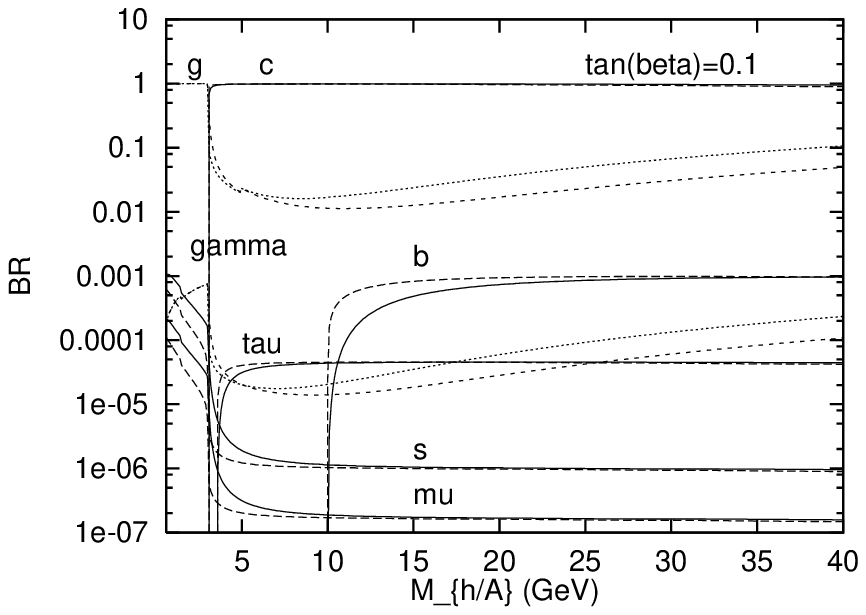}}
\vskip -0.15 in \relax\noindent\hskip 2.65in
                \relax{\includegraphics{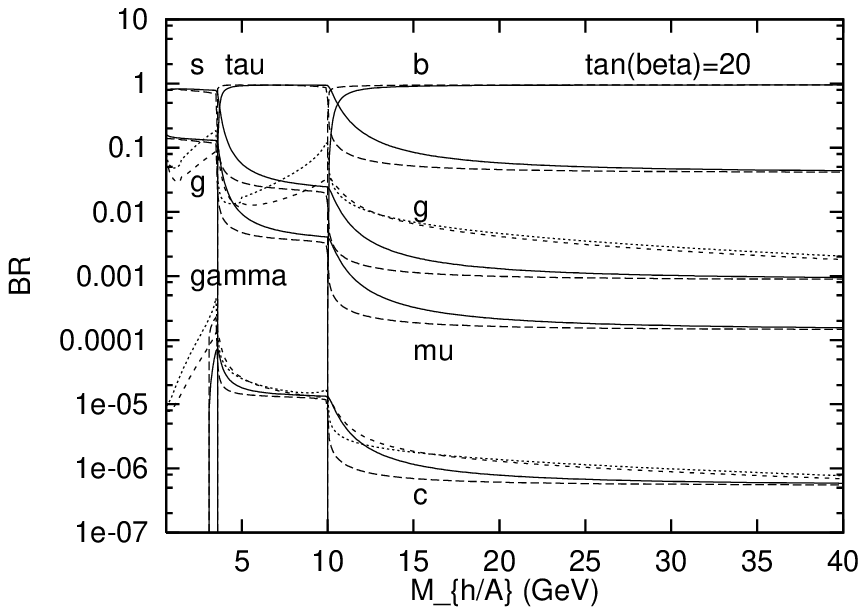}}
\vspace{-8.5ex}
\baselineskip 0.4cm
\caption{\sl  The branching ratio for a scalar boson decay
with $\alpha=\beta$ (solid lines for the fermionic modes) and
for a pseudoscalar one (dashed lines for the fermionic modes).
The corresponding decays into gluons and photons are denoted
by  short-dashed (scalar)  and the dotted (pseudoscalar) lines.
a) $\tb=0.1$, b) $\tb=20$.}
\label{fig:dec_aeqb}
\end{figure}

The preferred decay modes of Higgs bosons
depend on the parameters of the model.
For the condition $\alpha=\beta$ and masses of up to 40 GeV
the decay branching fractions of scalar and pseudoscalar Higgs bosons
are presented in Figs.~\ref{fig:dec_aeqb}a and b
for the two  choices $\tb=0.1$
and $\tb=20$.
They do not change significantly for smaller, respectively larger values
of $\tb $ and masses of up to $\sim $ 80 GeV.

The decay branching ratios are fairly similar for $h$ and $A$,
they differ only around the production thresholds
of the various fermion pairs.
In the case of  $\tb \gg$ 1 and masses above 4 GeV,
both $h$ and $A$ decay to almost 100$\% $ into $\tau $'s, or, once their 
threshold is passed, into beauty quarks.
For $\tb \ll $ 1 they decay almost exclusively into gluons and,
for $M_{h,A} > $ 2$m_c$, into charm quarks.
With increasing $M_{h,A}$ the decay into gluons 
rises again and reaches some 10$\% $ around 40 GeV.
 
\begin{figure}[ht]
\vskip 3.5in \relax\noindent\hskip 0.5in
                \relax{\includegraphics{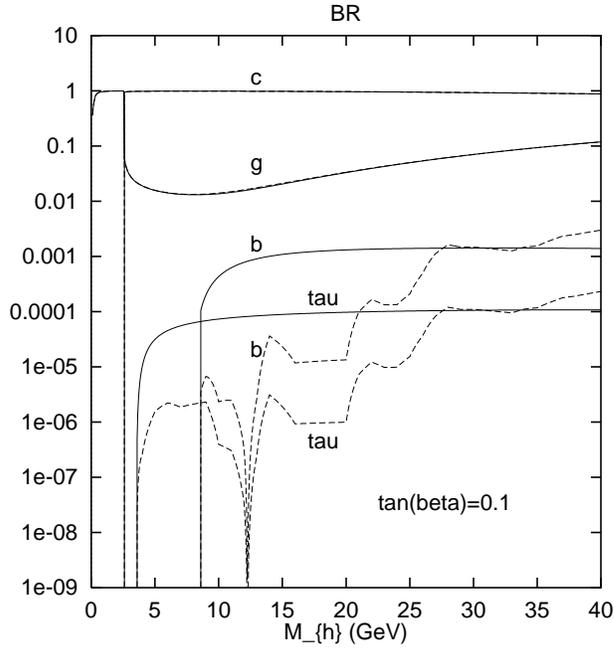}}
\vspace{-2.7ex}
\baselineskip 0.4cm
\caption{\sl  The branching ratio for a scalar boson decay
with the experimental limit on the  $\sin^2(\beta-\alpha)$ 
(dashed line) and with $\alpha=\beta$ (solid line) for  $\tb=0.1$.}
\label{fig:dec_sinL3}
\end{figure}

The branching fractions for decays of the scalar bosons
depend 
through $\sin^2(\beta-\alpha)$ also on the parameter $\alpha $.
For $\tb$ = 0.1 these fractions are compared in Fig.~\ref{fig:dec_sinL3}
for $\alpha=\beta$
and the maximum $\sin^2(\beta-\alpha)$ allowed by data \cite{bib-sinL3}.
No difference of relevance for experimental studies
is observed: the dominant decay modes
are hardly affected and only extremely suppressed branching
fractions exhibit some sensitivity.
Also for larger $\tb$ (not shown)
the leading decay modes are not affected by the value of 
$\sin ^2(\beta - \alpha )$.

% **********************************************************************

\section{Sensitivity to charged Higgs boson contribution}

Compared to the Standard Model an additional contribution involving
loops of charged Higgs bosons has to be included for the production
of a scalar $h$.
The relevant $hH^+H^-$ coupling in the general 2HDM 
\cite{hunter,dju,haba} is more complicated
than the Higgs couplings to fermions and gauge bosons.
It depends on the masses of both $M_h$ and $M_{H^{\pm}}$
and an additional parameter $\lambda_5$, 
remaining from the original Higgs potential:

\be
g_{hH^+H^-}={{M_h^2-\lambda_5v^2}\over{M_W^2}} {{\cos(\beta+\alpha)}
\over{\sin{2\beta}}}+{{2M_{\ch}^2-M_h^2}\over{2M_W^2}}\sin(\beta-\alpha)
\ee
where $\lambda_5$ is an arbitrary parameter and the vacuum expectation 
value: $v$=246 GeV (with a normalization, up to the sign, as for the gauge
boson in the SM, see Eq. 3).

In the following analysis we will assume that $\lambda_5$=0, 
which corresponds to the assumption of the strict symmetry of the Lagrangian 
under  
the scalar Higgs 
doublet transformation $\phi_1\ra -\phi_1$. 
In general, our results should be
correct for $|\lambda_5v^2|\ll M_h^2$.
Even for such small $\lambda_5$ it is still possible to have both
a decoupling 
and a non-decoupling of the heavy charged Higgs particle. 
In contrast to the belief stated eg. in \cite{dju}
that the $\Gamma(Z\ra h + \gamma)$ will be hardly sensitive to 
the charged Higgs particle loop, 
there are interesting parameter regions where one may expect 
to see such an effect.

Let us discuss this dependence in more detail.
We start by considering different values of $\sab$.
If $\sab$=0 we have the so called decoupling case,
as only the first term of $g_{hH^+H^-}$ (Eq.~6) contributes and 
therefore the overall contribution to the branching ratio 
due to the charged Higgs loop 
is given by 
\be
{{g_{hH^+H^-}}\over{M_{\ch}^2}}\propto 
               {{M_h^2}\over{M_{\ch}^2}} ({{1}\over{\tb}}-\tb),
\ee
leading to the negligible contribution for a very heavy charged 
Higgs boson.
(Here factors not relevant to our discussion are omitted.)
Note that the $W$ contribution, otherwise dominating 
the branching ratio for intermediate $\tb$, becomes negligible
for $\sab \sim$ 0 
and the effects due to the charged Higgs boson might be eventually seen 
if the mass of the charged Higgs is not too large, see below. 
For both very small and very large $\tb$ 
the $H^{\pm}$ may contribute with a strength that is
almost comparable to those from heavy quarks or W-bosons.
The difference in sign between the small and large $\tb$
scenarios may result in the constructive 
or destructive interference with bottom, or top quark, or W
contributions.
For $\tb$=1 the contribution from charged Higgs bosons disappears.

For $\sin(\beta-\alpha) \neq$0 and for $M_{\ch}\gg M_h$ 
the non-decoupling limit is obtained, and   
\be
{{g_{hH^+H^-}}\over{M_{\ch}^2}} \propto \sin(\beta-\alpha),
\ee
independent of the mass of Higgs bosons and $\tb$.

\begin{figure}[ht]
\vskip 4.in \relax\noindent\hskip -1.in
                \relax{\includegraphics{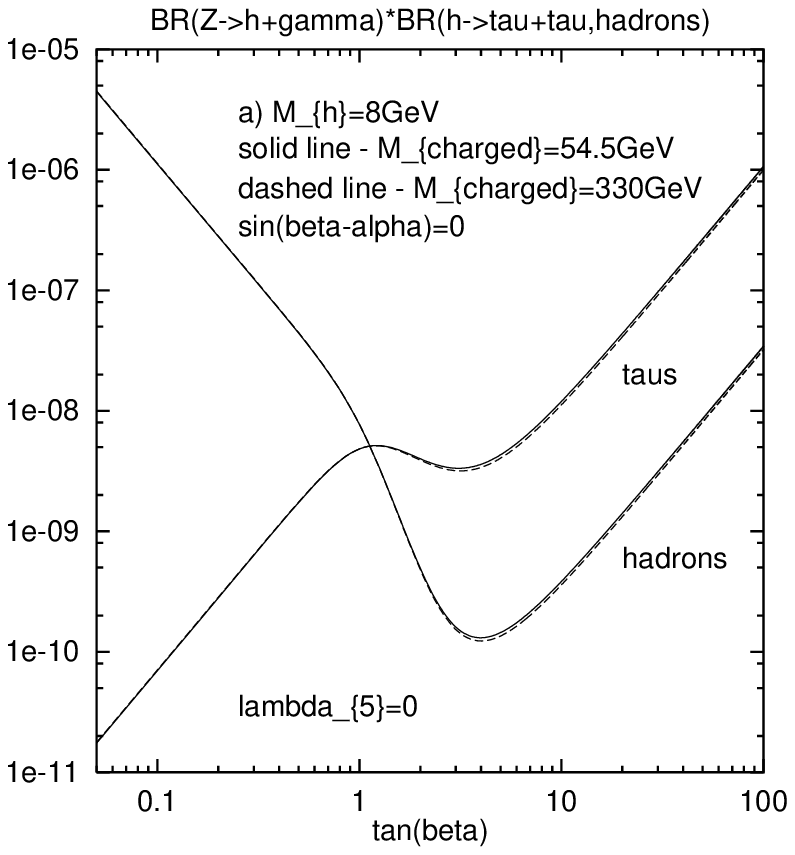}}
                  \relax\noindent\hskip 3.in
                \relax{\includegraphics{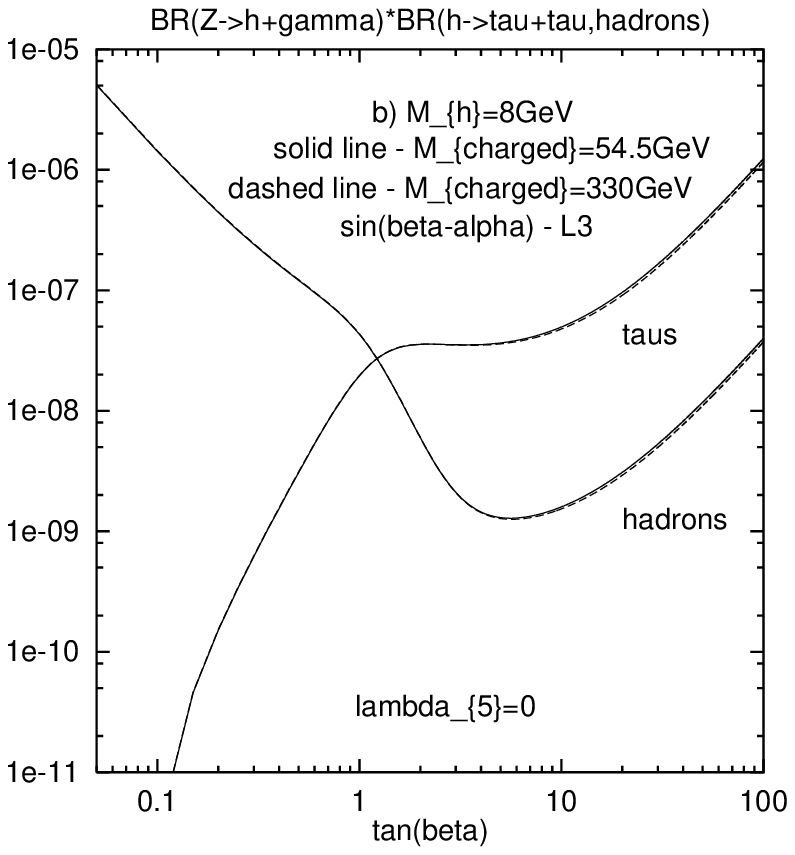}}
\vspace{-8.5ex}
\baselineskip 0.4cm
\caption{\sl   The branching ratio as a function of $\tb$
for a scalar boson decay with $M_h$=8 GeV with a) $\alpha=\beta$
  and b) the experimental limit on  $\sin^2(\beta-\alpha)$.
The mass of charged Higgs boson is equal to 54.5 and 330 GeV.
 }
\label{fig:prodbra1}
\end{figure}

\begin{figure}[ht]
\vskip 3.5in \relax\noindent\hskip -1.in
                \relax{\includegraphics{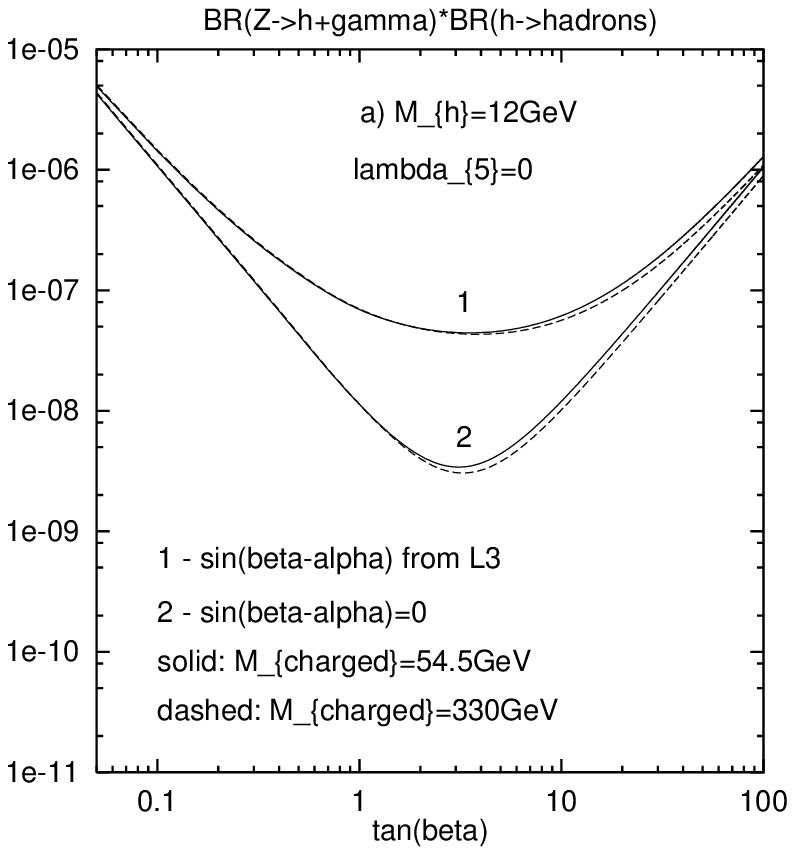}}
                  \relax\noindent\hskip 3.in
                \relax{\includegraphics{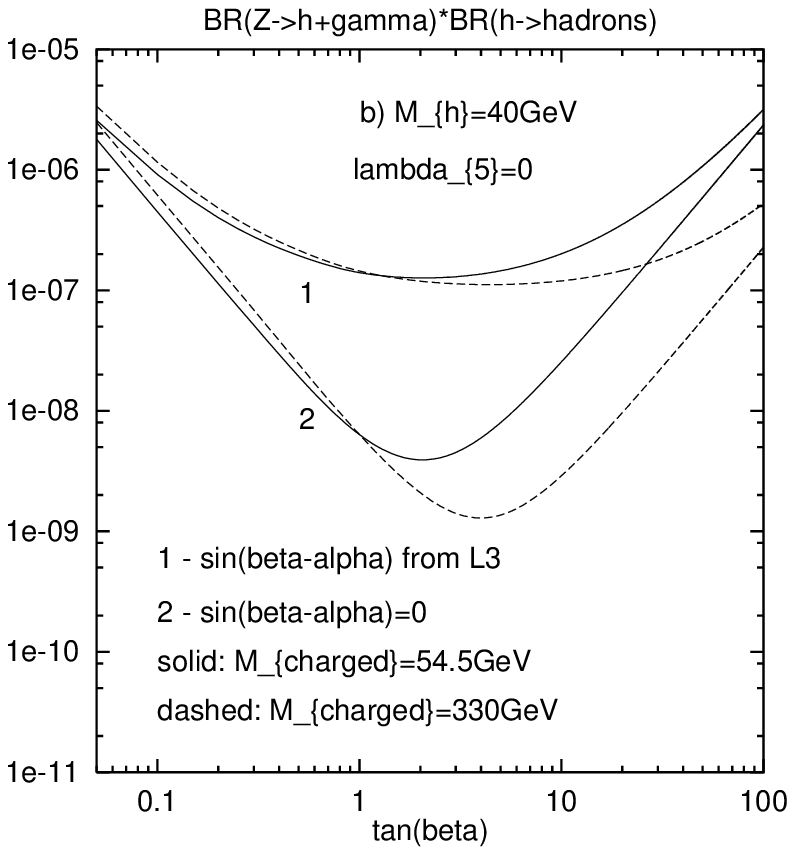}}
\vspace{-8.5ex}
\baselineskip 0.4cm
\caption{\sl   The branching ratio as a function of $\tb$
for a scalar boson decay with a) $M_h$=12 GeV
and b)  $M_h$=40 GeV .
The results obtained with the assumption
$\alpha=\beta$
and with  the experimental limit on  $\sin^2(\beta-\alpha)$ are plotted.
The mass of charged Higgs boson is equal to 54.5 and 330 GeV.
 }
\label{fig:prodbra2}
\end{figure}

The effect of charged Higgs bosons, assuming
$\lambda_5$=0, on
% In Figs.~\ref{fig:prodbra1} and \ref{fig:prodbra2} the product branching 
ratios 
$BR(Z\rightarrow h +\gamma)\cdot BR(h\rightarrow f\bar {f})$
is shown in Figs.~\ref{fig:prodbra1} and \ref{fig:prodbra2}
for the 'hadronic', i.e. the $qq+gg$, decay mode and the tau decay
channel.
The product branching ratios 
are presented as a function of $\tb $ 
for masses of the charged Higgs boson
of 54.5 and 330 GeV (which gives a similar result as for masses of 
1000 GeV or greater)
 and for masses of the scalar particle $h$
of 8, 12, and 40 GeV.
For $\sin^2(\beta-\alpha) $ = 0 
a smaller product branching ratio is observed, as expected. 

For a lower scalar mass of 8-12 GeV 
and for almost the whole range of $\tb$ 
the expected product branching ratios are  insensitive to the 
value of $M_{H^{\pm }}$.
However, with increasing mass $M_h$, the sensitivity to the mass
of the charged Higgs boson becomes more prominent
(cp. Eq. 7).
The contribution  of charged Higgs boson increases the $h$ 
production rate with diminishing  $M_{H^{\pm }}$ 
for $\tb \gg $ 1, but decreases it for $\tb \ll $ 1 (cp. Eq. 7).
(Note that the lower dashed curves (2) in Figs. 5a,b 
 can be treated as a bare fermionic contributions.)
The value of  $M_{H^{\pm }}$ affects the branching ratio stronger 
for large $\tb$ 
than for small $\tb$, where
the top loop interferes destructively with the charged Higgs boson
contribution.
At $\tb$=1 the charged Higgs boson does not contribute for 
$\sin^2(\beta-\alpha) $ = 0,
the point where its contribution disappears
(observe cross over points between solid and dashed lines in Fig. 5b)
  is shifted to slightly larger $\tb$ value for
the  $\sin^2(\beta-\alpha) $ = 0.25, the experimental limit 
for $M_h$=40 GeV~\cite{bib-L3}.

As we already mentioned above, 
the figures show a non-negligible dependence
on the parameter $\sin(\beta-\alpha)$
which governs the couplings $W^+W^-h$, $H^+H^-h$ and also (partly)
  $h f {\bar f}$.
In Figs.~\ref{fig:prodbra1} and \ref{fig:prodbra2} the branching ratio 
with  $\sin(\beta-\alpha)$=0 is compared to the
one accounting for the
experimental limit on $\sin(\beta-\alpha)$.
The difference due to $\sin^2(\beta-\alpha)$
is one to two orders of magnitude in the branching ratios for
intermediate values of $\tb$ but much less
for the extreme values of $\tb $ where it  has effects
at the 30 - 50 $\% $ level.

The effect due to charged Higgs boson loop 
should be larger for larger $M_h$ and will be studied elsewhere
for different assumptions on $\lambda_5$ \cite{my}.

\section{The experimental results}

\begin{figure}[p]
\center
\vskip 20.0cm \relax\noindent\hskip -8.3in
                   \relax{\includegraphics{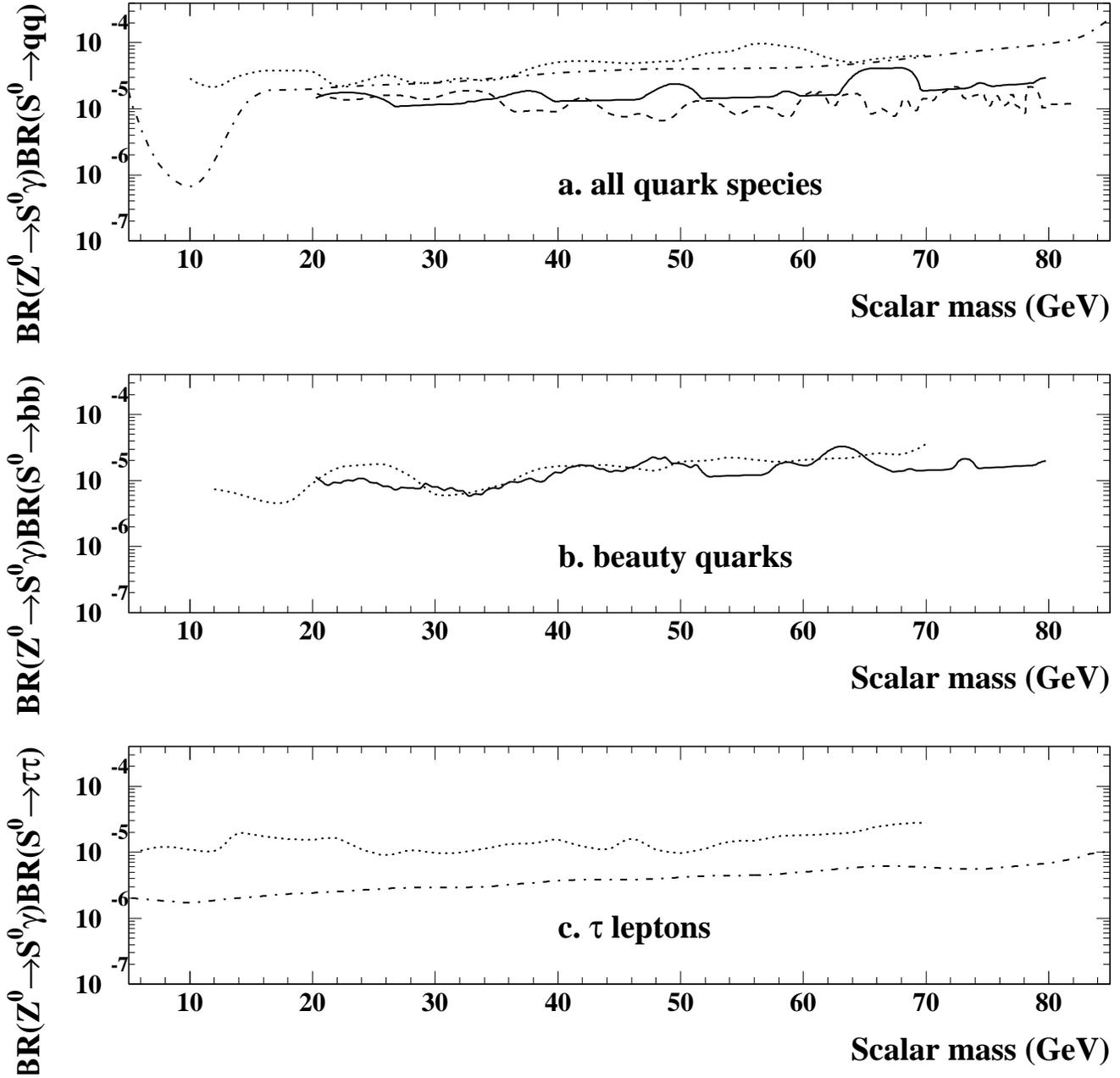}}
\vspace{-4.5cm}
\baselineskip 0.4cm
\caption{\sl Limits on the branching ratio $Z^0\rightarrow S+\gamma $
         from the various LEP experiments.
         Shown are the limits for the cases that the $S$ decays into
         any kind of quarks or gluons (a), into beauty quarks (b),
         or into $\tau $ pairs (c).
         ALEPH~\cite{bib-ALEPH}: dashed - dotted, 
         DELPHI~\cite{bib-DELPHI}: dotted, 
         L3~\cite{bib-L3}: dashed, 
         OPAL~\cite{bib-OPAL}: full.}
\label{fig:expt}
\end{figure}

All LEP experiments have searched for $Z$ decays into a scalar particle
$S$ and a photon.
Such particles would appear as a resonance peak of rather narrow
width over a background, which is mainly due to
photons emitted from the final state fermions. 
Results have been presented for the decay modes:
\begin{itemize}
\item $S\ \rightarrow \ \tau ^+ \tau ^-$ \cite{bib-ALEPH,bib-DELPHI}.
\item $S\ \rightarrow \ q \bar {q} $ without flavour tag
      \cite{bib-ALEPH,bib-DELPHI,bib-L3,bib-OPAL}.
      This can also be interpreted in terms of a decay into two
      gluons.
\item $S\ \rightarrow \ b \bar {b} $ \cite{bib-DELPHI,bib-OPAL}.
\end{itemize}
In addition decays into muons, electrons, neutrinos and photons
have been considered, but are of less interest 
in the context of Higgs searches in the 2HDM.
No single experiment has observed any significant structure,
the corresponding limits are shown in Fig.~\ref{fig:expt}.
~From this figure it becomes apparent that also a combination
of the results would not reveal any significant peak.
Thus, there exists no indication of a production of a 
Higgs boson in this process.

The LEP experiments considered explicitely only the
production of a scalar particle.
Since the angular distribution of $Z$ decays into a pseudoscalar
and the photon is identical,
the experimental limits, taking into account the different
normalization, can be directly applied also to the pseudoscalar Higgs boson
$A$.

The typical individual limits are 
$Br(Z\rightarrow S+\gamma )\cdot Br(S\rightarrow X) \sim 10^{-5}$ for
$X = q \bar {q}$ and $b \bar {b}$.
A notable exception is the result of \cite{bib-ALEPH} which sets limits
of less than $10^{-6}$ for $M_S\sim $ 10 GeV and 
$S\ \rightarrow \ q \bar {q} $.
For $X = \tau ^+ \tau ^-$ limits have been set between 
2$\cdot 10^{-6}$ at $M_S \sim $ 5 - 20 GeV and 
$10^{-5}$ at $M_S \sim $ 85 GeV.
Without more detailed information, for example about the mass dependent
backgrounds, data yields and efficiencies, it is impossible 
to combine the results from the various experiments in a rigorous manner.
Generally one expects the limits to improve by some factor $\sqrt {2}$ - 2.
In the absence of this 
detailed information we will consider the most restrictive
limit from all experiments.
This is justified because of the absence of a consistent indication 
of a signal.
In general, though not necessarily everywhere, this approach should be
conservative.

For $\tb \ll $ 1 the $h$ and $A$ decay into charm quarks and gluons to
almost 100$\% $.
Limits on both of these decays are not explicitely provided by the 
experiments.
However, since charm as well as gluon jets are 
rather similar to those of other flavours,
no significant change of the experimental efficiency compared
to the study of inclusive quark decays should be 
expected~\footnote{
The ALEPH collaboration has explicitely studied $S\ \rightarrow gg $
and obtains limits which are almost identical 
to those for $S$ decays into inclusive quark flavours
\cite{bib-ALEPH}.}.
In considering the low $\tb $ region,
we therefore apply the limits from inclusive decays into quarks (and gluons).

\section{Results }

\begin{figure}[hb]
\vskip 3.5in \relax\noindent\hskip -1.in
                \relax{\includegraphics{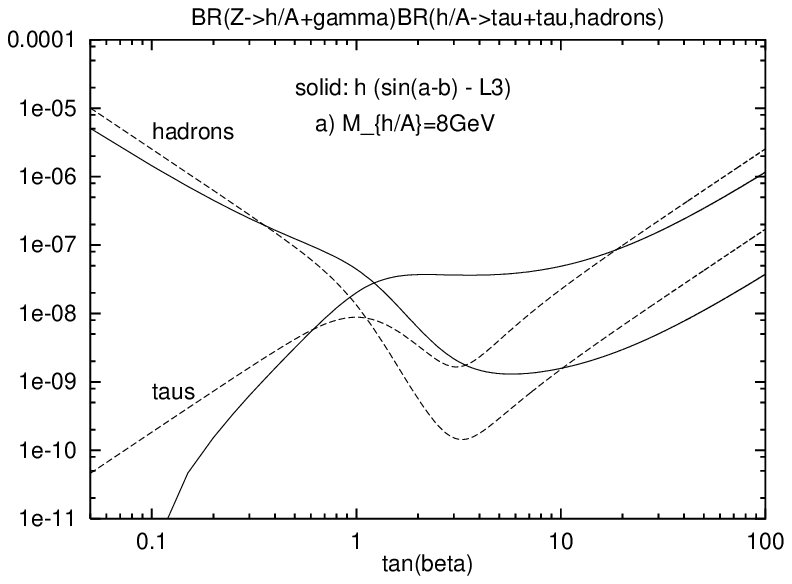}}
%\vskip -1.8cm
                  \relax\noindent\hskip 3.2in
                \relax{\includegraphics{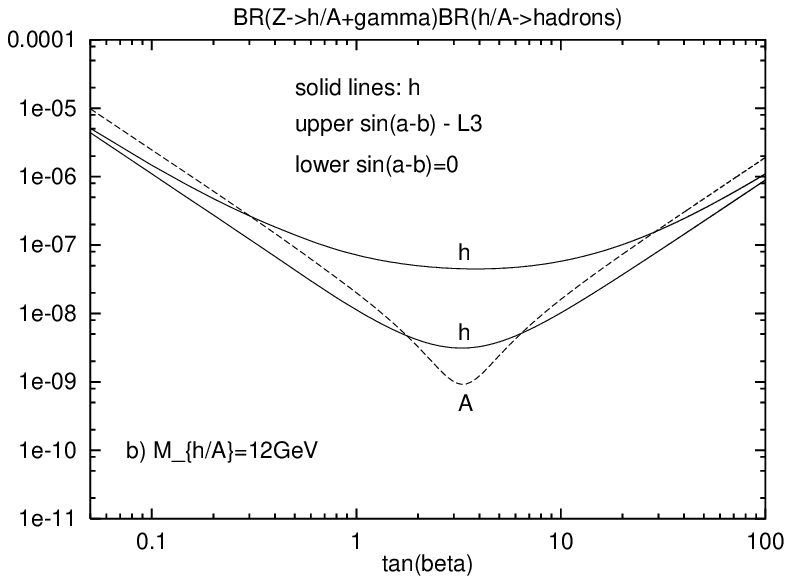}}
\vskip 3.0 in \relax\noindent\hskip -1.in
                \relax{\includegraphics{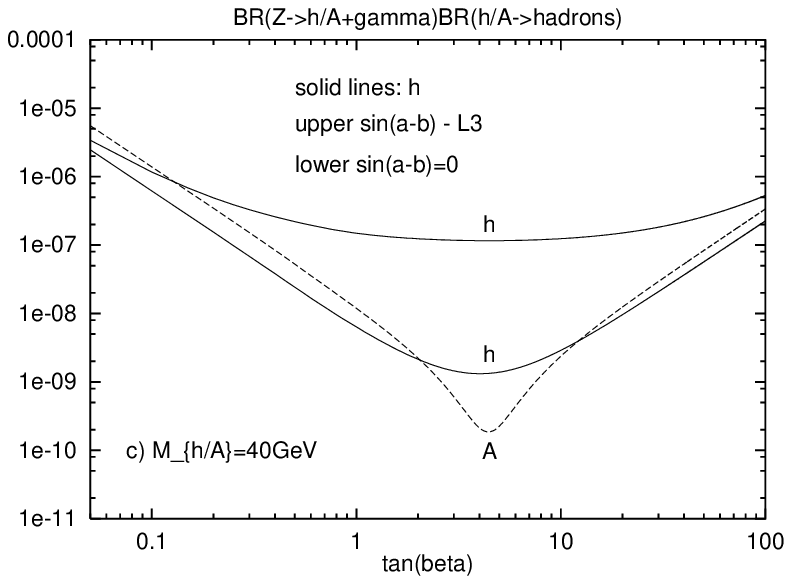}}
                  \relax\noindent\hskip 3.2 in
                 \relax{\includegraphics{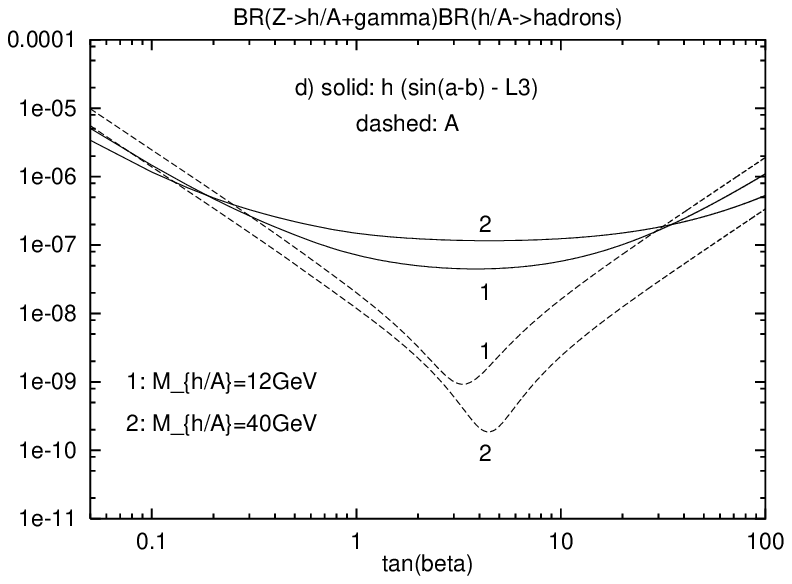}}
\vspace{-8.5ex}
\baselineskip 0.4cm
\caption{\sl  a)
The branching ratio for a pseudoscalar boson (dashed line)
compared to the scalar case (solid line) for 
 $M_{h(A)}$=a) 8, b) 12, and c) 40 GeV, respectively.
The $h$ curves take into account the experimental limits on
$\sin ^2(\beta -\alpha)$ and assume
$M_{H^{\pm}} $=330 GeV.
The $X=qq+gg$ is denoted by 'hadrons', while $X=\tau\tau$ is described by 
'taus'.
In d) a comparison is made for two masses 12 and 40 GeV.}
\label{fig:exclusion}
\end{figure}

The product branching ratios 
$Br(Z\rightarrow h(A) +\gamma) \cdot Br(h(A)\rightarrow X)$
% with the experimental limit on 
are plotted in Figs.4,5 discussed above for the scalar case 
and in Figs.~\ref{fig:exclusion}a,b,c,d as a function of 
$\tb$ for $h$ and $A$ masses of 8, 12 and 40 GeV.
Because the experimental sensitivity to other decay modes 
is rather limited,
only the $qq+gg$ decay mode denoted 'hadronic'
and the decay into $\tau$'s (for $M_{h(A)}$=8 GeV) are considered.
The experimental limits on $\sin (\beta - \alpha )$ and on the
mass of the charged Higgs particle 
are taken into account.
  
These product branching ratios agree within up to about a factor two 
for scalar and pseudoscalar Higgs bosons and
values of $\tb$ of less than $\sim $ 0.2 and 
larger than $\sim $ 50. 
They differ drastically for intermediate values
of $\tb$, where the pseudoscalar production rate can be lower 
by some two orders of magnitude. 
This difference is mainly due to
the additional contribution of $W$ loops for the $h$ production (see for 
example Figs. 7b,c).

One sees that in general the experimental limits 
on the product branching ratio 
$Br(Z\rightarrow h(A) + \gamma) \cdot Br(h(A)\rightarrow X)$ of
$\sim$ 10$^{-5}-10^{-6}$
are significantly above the
expected rates for a wide range of $\tb $ values.
An exception are the extremely high and low values of $\tb $.
Here
the data impose additional constraints on the 2HDM.
This is especially true in the mass region
$\sim$ 10 GeV, where an experimental sensitivity of 
below $10^{-6}$ is reached. 
Limits on $\tb $ as a function of the $h$ and $A$ masses are shown in
Fig. 8.
The constraints in the two extreme regions of $\tb $ can be summarized 
as follows.

\begin{itemize}

\item In the region of $\tb \ll $ 1 the product branching ratio 
      $Br(Z\rightarrow h(A)+\gamma )\cdot Br(h(A)\rightarrow X)$ is 
      larger than $10^{-6}$ 
      for masses of up to 40 GeV.
      Here the non-observation of associated $h(A)+\gamma $ decays leads to new
      constraints. 
      Unfortunately only around 
      10 GeV the data exclude values of $\tan \beta $ that are not disfavoured
      by theoretical arguments.

\item Also in the region of high $\tb $, ${\cal O}$(100), the data limit the
      $\tb $ range.
      It is constrained to be smaller than 75 (55) (for $M_{h}(M_A)=$ 10 GeV) 
and
      smaller than ${\cal O}$(300) (for $M_{h(A)}=$ 35 GeV).
      These constraints are  around 10 GeV more stringent
      than the limits from todays (g-2)$_{\mu }$ data \footnote{
      Those are expected to be improved soon by the 
      E821 experiment at BNL~\cite{e821}.}.
      They are, however, less restrictive than the constraints from the
      Yukawa process  
      \footnote{As mentioned in the introduction,
      the experimental results for the Yukawa process
      have as yet been presented
      only for pseudoscalars $A$.
      However, an interpretation in terms of a potential scalar production 
would
      yield stronger limits on $\tb $.}.
\end{itemize}

The limits on $\tb$ as a function of $M_h$ and $M_A$ 
were obtained for $\lambda_5$=0 and 
a charged Higgs mass 330 GeV and for comparison also for mass 54.5 GeV, but, 
as long as it is above ~200 GeV 
the limits will change only marginally. 
The dependence (for $h$ only) 
on the assumption on the $\sin^2(\beta-\alpha)$
on the obtained limits is  weak,
and the exclusion plot in Fig. 8 
corresponds to  the tightest limit on $\tb$ corresponding to
the experimental limits on $\sin^2(\beta-\alpha)$.
Assuming $\alpha=\beta$ the limit will
be weaker, being 
 shifted up and down  by approximately factor of 1.4 for the mass of 40 GeV,
 for lower masses the change will be much smaller.
 
Also shown in Fig.~8 is a dependence on the  mass of the charged Higgs boson
 for the $h$ and larger $M_h$ values 
(a difference by the solid curves ``1'' ($M_{H^{\pm}}$=54.5 GeV)  and ``2''
 (330 GeV)).  
% ========================================================================

\section{Conclusion and outlook}

The one-loop result to the process
$Z{\rightarrow}h+\gamma$ in  the general Two Higgs Doublet Model (Model II)
is compared to  the experimental limits from LEP, 
which is of the order 
$BR(Z\rightarrow h +\gamma) \cdot BR(h\rightarrow X)$
$\sim 10^{-6}-10^{-5}$.
Taking into account the existing limits on $\sin^2(\beta-\alpha)$
we analysed the light mass of neutral scalar Higgs bosons scenarios
with large and small $\tb$.
We find that the process constrains the parameter space to 
$\tb$ between 0.15 and 70 
for masses $M_h \ \sim $ 10 GeV.

We  studied the dependence of the scalar production yield 
a $\sin^2(\beta-\alpha)$ and the mass of the charged Higgs boson.
The parameter $\sin^2(\beta-\alpha)$ induces
 dependence for intermediate $\tb $,
but affects only mildly the product branching ratios at extreme
values of $\tb $. The dependence on the charged Higgs mass, 
in the limits of 54.5 to 330 GeV becomes stronger with higher mass
$M_h$.

The product branching ratios for the associated 
production of a pseudoscalar $A$ and a photon
is similar to the one for scalars for $\tb$ below 0.2 and $\tb$ 
above 50. Thus similar limits to those for the scalar $h$ can be derived. 
They differ drastically for intermediate values
of $\tb$, where the pseudoscalar production rate is much lower,
than for scalars because of the strong $W$ contribution in the latter case.

For a large parameter space of the 2HDM the data have no sensitivity to the
expected yields.
Only for extremely high or low values of $\tb $ some constraints can be 
derived. 
The large $\tb$ region of the 2HDM 
can be constrained by the data.
These limits are stronger than those from the present 
$g-2$ data  for muons for both a light scalar and a light 
pseudoscalar Higgs boson. 
For the light pseudoscalar scenario the existing data
from the Yukawa process at LEP lead to stronger limits, 
but for mass around 10 GeV
the $Z\ra A + \gamma$ decay becomes competitive.

Constraints on the 2HDM model can also be obtained for the low values 
of $\tb$ for both scalar (similar remarks as above 
for the large $\tb$ case hold here as well) and pseudoscalar production.
Also these limits are of interest, although they just touch the 
region of $\tb \ll $  1, which is required by perturbative 
calculations. 

To summarize, the process discussed here leads to constraints of 
the parameters of the 2HDM for very large and very low $\tb$ for 
both scalar and pseudoscalar production. The one-loop
calculation applied here may be improved in the future by taking into account 
higher order corrections. 

Finally let us consider possible experimental improvements.
Although data taking at the $Z$ has been completed, some improvements
may be expected from the data since not the whole statistics 
has been used up to now for the various analyses
and improvements seem possible.
Only one experiment has fully exploited the $\tau \tau \gamma $
channel, the low mass region $\le $ 20 GeV has also not been addressed
by all experiments and finally most experiments have improved their 
beauty tagging compared to what has been published.
Assuming in addition
a proper combination of the final data it may be possible to gain
some factor 2-4 in sensitivity.
This would imply a sensitivity to branching ratios of some 10$^{-6}$.
The drastically lower cross section at the high energies of 160 - 200 GeV
of LEP
and also the increased background from
initial state photons above the $Z$ pole,
renders it unlikely that a sensitivity close to
the expected yields within the 2HDM can be reached.
On the other hand, higher masses can be reached which in itself makes it
important to consider this process.
A first look~\cite{bib-DELPHI_LEP2}, however, did not reveal any new
particle production.
The interpretation of experimental results require a more general theoretical 
analysis
which includes not only the production of on-shell $Z$ decays.

The high luminosities which are envisaged at a new linear $e^+e^-$ collider
or $\mu ^+ \mu ^-$ collider
may allow some sensitivity to the associated $h(A) +\gamma $ production.
However a detailed experimental study is still missing.

\vspace{0.5cm}

{\large \bf Acknowledgments}

\vspace{0.25cm}

One of us (MK) is  grateful 
the Physics Faculty at 
Dortmund University for the kind hospitality during her visit
when the project started.
She is also indebted to Theory Group at DESY for the
support and warm hospitality.
The useful discussions with A. Djouadi and P. Zerwas are   
acknowledged. 
MK is grateful to M. Peskin  for the discussion on the angular
dependence of the final photon and on the mass limits for a light Higgs boson
from the Wilczek process \cite{wilk}c and to 
T. Shimada  pointing us the Ref.\cite{shimada}.
She is indebted to
H. Haber and M. Carena for very important  suggestions
and  hospitality
during her stay at Santa Cruz and FERMILAB, when the 
paper was finalized.
MK acknowledges the critical comment by Sally Dawson about the earlier
version of figures.
J\.Z is very grateful to M. Staszel and A. Zembrzuski for 
discussions, help and assistance with preparing this article.
We also thank our colleagues Terry Medcalf (ALEPH) and Joachim Mnich
(L3) for providing us with the numerical values of the respective
experimental limits.
 
This work was 
supported partially by Polish Committee 
for Scientific Research, grant N\b{o}. 
2P03B18209 and 2P03B01414 and by US-Poland Maria Sk\l odowska - Curie
Joint Fund II (MEN/DOE --96--264).

% **********************************************************************

\begin{figure}[ht]
\vskip 5.in \relax\noindent\hskip -0.50in
                \relax{\includegraphics{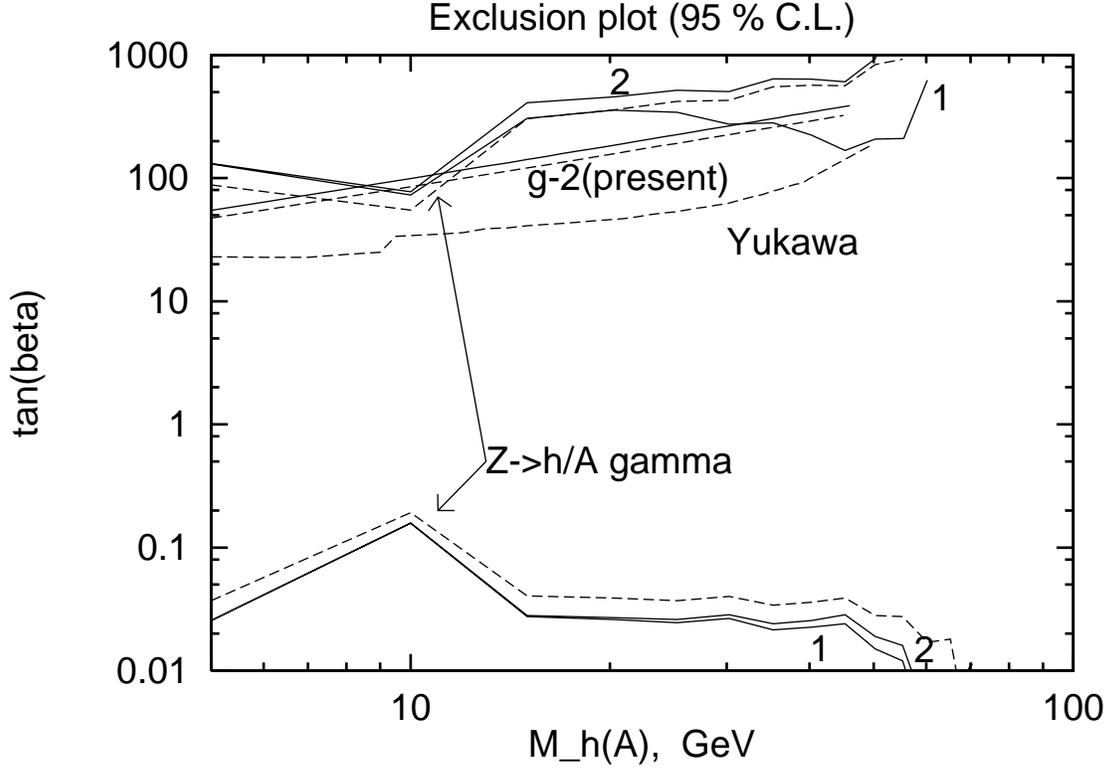}}
\vspace{-10.5ex}
\baselineskip 0.4cm
\caption{\sl  The exclusion plot (95\% C. L.) for the $\tb$
versus mass of the scalar (with experimental limits on $\sin(\beta-\alpha)$,
solid  line)
or the pseudoscalar( dashed line) obtained from the
data on  $Br(Z\ra h(A) + \gamma)$ for  the hadronic final state
with an exception of  
 the
lowest mass uses the $tau$-channel,
data from OPAL, L3 and ALEPH (below 20 GeV). For scalar production
 two masses for the charged
Higgs boson  were used:
1 -- 54.5 GeV and 2 -- 330 GeV. 
For comparison  exclusion based on 
 ALEPH data from Yukawa process and present $g$-2 for muon
measurement is  shown.
The area above upper and below lower curves is excluded.
}
\end{figure}

\end{document}